\documentclass[pra,showpacs,twocolumn]{revtex4}
\usepackage{amsfonts}

\usepackage{psfrag}
\usepackage{graphicx}
\usepackage{bbold}

\begin{document}

\title{Entanglement and local extremes at an infinite-order quantum phase
transition}
\author{C. C. \surname{Rulli}}
\email{rulli@if.uff.br}
\affiliation{Instituto de F\'{\i}sica, Universidade Federal Fluminense, Av. Gal. Milton
Tavares de Souza s/n, Gragoat\'a, 24210-346, Niter\'oi, RJ, Brazil.}
\author{M. S. \surname{Sarandy}}
\email{msarandy@if.uff.br}
\affiliation{Instituto de F\'{\i}sica, Universidade Federal Fluminense, Av. Gal. Milton
Tavares de Souza s/n, Gragoat\'a, 24210-346, Niter\'oi, RJ, Brazil.}
\date{\today }

\begin{abstract}
The characterization of an infinite-order quantum phase transition (QPT) by
entanglement measures is analyzed. To this aim, we consider two closely
related solvable spin-1/2 chains, namely, the Ashkin-Teller and the
staggered XXZ models. These systems display a distinct pattern of
eigenstates but exhibit the same thermodynamics, i.e. the same energy
spectrum. By performing exact diagonalization, we investigate the behavior
of pairwise and block entanglement in the ground state of both models. In
contrast with the XXZ chain, we show that pairwise entanglement fails in the
characterization of the infinite-order QPT in the Ashkin-Teller model,
although it can be achieved by analyzing the distance of the pair state from
the separability boundary. Concerning block entanglement, we show that 
both XXZ and Ashkin-Teller models exhibit identical von Neumann entropies as long as a
suitable choice of blocks is performed. Entanglement entropy is then shown to be able to 
identify the quantum phase diagram, even though its local extremes (either
maximum or minimum) may also appear in the absence of any infinite-order QPT. 
\end{abstract}


\maketitle


\section{Introduction}

The behavior of entanglement in many-body systems has attracted great
attention in recent years due to its promising potential to realizing
quantum information tasks~\cite{Kane:98,DiVincenzo:00,Bose:03} as well as
its relationship with quantum critical phenomena~\cite{Osterloh:02}. In
particular, it has been observed that entanglement can identify and
characterize a quantum phase transition (QPT)~\cite%
{Sachdev:book,Continentino:book}. QPTs are associated with critical changes
in the ground state of a quantum system due to level crossings in its energy
spectrum, occurring at low temperatures $T$ (effectively $T=0$).
Specifically, a first-order QPT is characterized by a finite discontinuity
in the first derivative of the ground state energy. Similarly, a
second-order QPT -- or a continuous QPT -- is characterized by the existence
of an infinite correlation length and a power-decay of correlations, which
is often manifested by a finite discontinuity or divergence in the second
derivative of the ground state energy, assuming the first derivative is
continuous. A more subtle class is the so-called infinite-order QPTs, for
which any finite-order derivative of the ground state energy is a continuous
function of the relevant parameters. Prominent examples of such QPTs are
provided by the metal-insulator point in the fermionic Hubbard model and the
SU(2) (antiferromagnetic) Heisenberg point in the XXZ spin-1/2 chain.

In recent years, it has been noticed that the behavior of entanglement (or its
derivatives) in the ground state of a many-body system undergoing a first-order 
or a continuous QPT exhibits a non-analiticity. Indeed, in the case of first-order QPTs,
discontinuities in the ground state entanglement were shown to detect the
QPT \cite{Bose:02,Alcaraz:03,JVidal:04}. For the case of second-order QPTs,
the critical point is found to be associated with a singularity in the
derivative of the ground state entanglement, as first illustrated for the
transverse field Ising chain in Ref.~\cite{Osterloh:02}, and generalized in
Refs.~\cite{Osborne:02,JVidal:04a,Huang:04} (see also Refs.~\cite%
{Verstraete:04a,Barnum:04,Somma:04,Vidal:03,Vidal:04} for an analysis in
terms of other entanglement measures). The behavior of entanglement for
first-order and second-order QPTs have also been discussed in general
grounds in Refs.~\cite{Wu:04,Wu:06}. For infinite-order QPTs, although no
independent-model analysis is available, entanglement has been found to
exhibit a local extreme (either maximum or minimum) at the quantum critical
point (QCP). This has indeed been shown for the spin-1/2 XXZ chain~\cite%
{Gu:03,Chen:06,Gu:07,Alcaraz:08} and for the fermionic Hubbard model~\cite{Gu:04}.
Whether or not this is a general property of an entanglement measure for a
convenient partition of the system remains unresolved.

In order to make progress on this matter and on the general properties of an
infinite-order QPT, we will consider in this paper the characterization of
entanglement into two closely related solvable spin-1/2 chains, namely, the 
Ashkin-Teller and the staggered XXZ models. The Ashkin-Teller model has been
introduced as a generalization of the Ising spin-1/2 model to investigate
the statitiscs of two-dimensional lattices with four-state interacting sites~%
\cite{Ashkin:43}. Since then, it has attracted a great deal of attention due
to its motivations in a wide range of research fields. First, both its
classical and quantum versions exhibit a rich phase diagram~\cite%
{Kohmoto:81,Rittenberg:87}, which makes the model a prototype for the
investigation of phase transitions and critical phenomena. For instance, it
has recently been shown that the quantum Ashkin-Teller model in
one-dimension exhibits an example of disorder rounding of a first order
quantum phase transition into a continuous phase transition~\cite{Goswami:08}, 
which may find applications to numerous complex strongly correlated
systems. Second, it has been experimentally realized by magnetic compounds
formed by layers of atoms adsorbed on clean surfaces, e.g., Selenium
adsorbed on Ni(100) surface~\cite{Bak:85}. Moreover, 
the universality properties Ashkin-Teller model may also be related to
many other intereseting applications, such as nonabelian anyons models~\cite%
{Gils:09}, orbital current loops in CuO2-plaquettes of high-Tc cuprates~\cite%
{Gronsleth:08}, and elastic response of DNA molecules~\cite{Chang:08}. 
Remarkably, the Ashkin-Teller model and the XXZ chain display a
distinct pattern of eigenstates but exhibit the same thermodynamics, i.e.
the same energy spectrum. In particular, the same quantum phase diagram
applies to both models, although entanglement can be distinct in each case.
Our aim is then analyze this quantum phase diagram and investigate to what 
extent entanglement measures will agree on
the description of the infinite-order QCP. Indeed, as we will
see, in contrast with the XXZ chain, pairwise entanglement between nearest
neighbors in the Ashkin-Teller model is not able to identify the QPT. By
focusing on block entanglement, we will then show how to reconcile the
behavior of entanglement in both cases. Moreover, we will show that local extremes 
may also occur in the absence of any infinite-order QPT, indicating that further
analysis of the critical behavior is demanded in a scenario where
entanglement displays a local maximum or minimum driven by a relevant parameter.


\section{The Ashkin-Teller model and its map into the staggered XXZ chain}


\label{map}

Let us begin by introducing the quantum Ashkin-Teller model in
one-dimension, whose Hamiltonian for a chain with $M$ sites is given by 
\begin{eqnarray}
&&H_{AT} =-J\sum_{j=1}^{M}\left( \sigma_{j}^{x}+\tau_{j}^{x} + \Delta
\sigma_{j}^{x} \tau_{j}^{x}\right)  \nonumber \\
&&\hspace{-0.5cm}-J\,\beta \sum_{j=1}^{M}\left( \sigma_{j}^{z}\sigma_{j+1}^{z}
+ \tau_{j}^{z}\tau_{j+1}^{z} + \Delta \sigma_{j}^{z}\sigma_{j+1}^{z}
\tau_{j}^{z}\tau_{j+1}^{z}\right),  \label{at}
\end{eqnarray}
where $\sigma_j^\alpha$ and $\tau_j^\alpha$ $(\alpha = x,y,z)$ are
independent Pauli spin-1/2 operators, 
$J$ is the exchange coupling constant, 
$\Delta$ and $\beta$ are (dimensionless) parameters, 
and periodic boundary conditions (PBC) are adopted, i.e., $%
\sigma^\alpha_{M+1} = \sigma^\alpha_{1}$ and $\tau^\alpha_{M+1} =
\tau^\alpha_{1}$ ($\alpha = x,y,z$). The Ashkin-Teller model is $Z_2 \otimes
Z_2$ symmetric, with the Hamiltonian commuting with the parity operators 
\begin{equation}
\mathcal{P}_1 = \prod_{j=1}^{M} \sigma_j^x \hspace{1cm} {\text{and}} \hspace{%
1cm} \mathcal{P}_2 = \prod_{j=1}^{M} \tau_j^x .  \label{parity-at}
\end{equation}
Therefore, the eigenspace of $H_{AT}$ can be decomposed into four disjoint
sectors labelled by the eigenvalues of $\mathcal{P}_1$ and $\mathcal{P}_2$,
namely, $Q=0$ $(\mathcal{P}_1 = + 1, \mathcal{P}_2 = + 1)$, $Q=1$ $(\mathcal{%
P}_1 = + 1, \mathcal{P}_2 = - 1)$, $Q=2$ $(\mathcal{P}_1 = -1, \mathcal{P}_2
= - 1)$, and $Q=3$ $(\mathcal{P}_1 = -1, \mathcal{P}_2 = + 1)$. By the
symmetry of $H_{AT}$ under the interchange $\sigma^\alpha \leftrightarrow
\tau^\alpha$, the sectors $Q=1$ and $Q=3$ are degenerate. Moreover, we
observe that the ground state belongs to the sector $Q=0$.

\begin{figure}[ht]
\centering {\includegraphics[angle=0,scale=0.38]{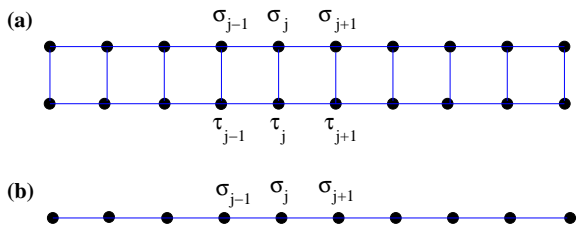}}
\caption{(Color online) (a) Ashkin-Teller spin-1/2 chain. The lattice is
composed by two independent spin-1/2 particles \textit{per} site $j$
described by Pauli operators $\{\protect\sigma^ \protect\alpha_j, \protect%
\tau^ \protect\alpha_j\}$. (b) XXZ spin-1/2 chain. The lattice is composed
by one spin-1/2 particle \textit{per} site $j$ described by Pauli operators $%
\{\protect\sigma^ \protect\alpha_j\}$. }
\label{f1}
\end{figure}

In order to map the Ashkin-Teller model into the staggered XXZ chain, we
consider two sets of $2M$ link variables $\{\eta_j, \gamma_j |
j=1,\cdots,2M\}$, which are defined by 
\begin{eqnarray}
&&\eta_{2j-1} = \sigma_j^x , \hspace{1cm} \gamma_{2j-1} = \tau_j^x, 
\nonumber \\
&&\eta_{2j} = \sigma_j^z \sigma_{j+1}^z, \hspace{1cm} \gamma_{2j} = \tau_j^z
\tau_{j+1}^z .  \label{lv}
\end{eqnarray}
These variables satisfy the conditions 
\begin{equation}
\eta_j^2 = \mathbb{1} , \hspace{1cm} \gamma_j^2 = \mathbb{1} \hspace{0.5cm}
(j = 1, \cdots, 2M) .  \label{linkcond}
\end{equation}
Moreover, they obey the algebra 
\begin{equation}
\left[ \eta_j , \gamma_k \right] = 0 , \,\,\, \left[ \eta_j , \eta_k \right]
= 0 , \,\,\, \left[ \gamma_j , \gamma_k \right] = 0 ,  \label{linkalg1}
\end{equation}
for $|j-k| \ne 1$ and $(j,k) \ne (1,2M)$ and $(j,k) \ne (2M,1)$, while 
\begin{equation}
\left[ \eta_j , \gamma_k \right] = 0 ,\,\,\, \left\{ \eta_j , \eta_k
\right\} = 0 , \,\,\, \left\{ \gamma_j , \gamma_k \right\} = 0 ,
\label{linkalg2}
\end{equation}
for $|j-k| = 1$ or $(j,k) = (1,2M)$ or $(j,k) = (2M,1)$. However, note
that, since PBC are adopted, the link variables are not completely
independent, turning out to obey the constraints 
\begin{equation}
\prod_{j=1}^{M} \eta_{2j} = \prod_{j=1}^{M} \gamma_{2j} = \mathbb{1}.
\label{cat}
\end{equation}
Moreover, by writing out $H_{AT}$ in terms of the link variables, we obtain 
\begin{eqnarray}
&&H_{AT} =-\sum_{j=1}^{2M}J\left( \eta_{2j-1} + \gamma_{2j-1} + \Delta
\eta_{2j-1} \gamma_{2j-1}\right)  \nonumber \\
&&\hspace{-0.5cm}-J\,\beta \sum_{j=1}^{2M}\left( \eta_{2j} + \gamma_{2j} +
\Delta \eta_{2j} \gamma_{2j} \right).  \label{at2}
\end{eqnarray}
The equivalence between the Ashkin-Teller model and
the XXZ chain can be established by considering the staggered spin-1/2 XXZ chain with $2M$
sites, whose Hamiltonian reads 
\begin{eqnarray}
&&H_{XXZ} =-\sum_{j=1}^{M}J\left[ \sigma_{2j-1}^{x}\sigma_{2j}^{x} +
\sigma_{2j-1}^{y}\sigma_{2j}^{y} -\Delta \sigma_{2j-1}^{z}\sigma_{2j}^{z}%
\right]  \nonumber \\
&&-J\beta \sum_{j=1}^{M}\left[ \sigma_{2j}^{x}\sigma_{2j+1}^{x} +
\sigma_{2j}^{y}\sigma_{2j+1}^{y} -\Delta \sigma_{2j}^{z}\sigma_{2j+1}^{z}%
\right],  \label{xxz}
\end{eqnarray}
where,
as before, $J$ is the exchange coupling constant,
$\Delta$ and $\beta$ are the anisotropy and staggering (dimensionless) parameters, respectively, 
and PBC are adopted, i.e., $\sigma^\alpha_{2M+1} =\sigma^\alpha_{1}$ ($\alpha = x,y,z$). 
Note that, differently from the
Ashkin-Teller model, the XXZ chain has one spin-1/2 particle \textit{per}
site. The XXZ model is U(1) invariant, with the Hamiltonian commuting with
the total spin operator $S^z = \sum_{j=1}^{2M} {\sigma}_j^z$, i.e., 
\begin{equation}
[ H_{XXZ}, S^z ] = 0.  \label{u1}
\end{equation}
Therefore, the eigenspace of $H_{XXZ}$ can be decomposed into $2M+1$
disjoint sectors labelled by their corresponding magnetization quantum
number $n = M - r$, with $r=0,1,...,2M$ denoting the number of spins
reversed from the state with all spins down. We observe that the ground
state belongs to the sector $n=0$. A schematic view of the XXZ and
Ashkin-Teller chains is shown in Fig.~\ref{f1}. 

As defined for the Ashkin-Teller model, we introduce two sets of link
variables, $\{\eta_j, \gamma_j | j=1,\cdots,2M\}$, which are given by 
\begin{eqnarray}
&&\eta_{2j-1} = \sigma_{2j-1}^x \sigma_{2j}^x, \hspace{1cm} \gamma_{2j-1} =
\sigma_{2j-1}^y \sigma_{2j}^y,  \nonumber \\
&&\eta_{2j} = \sigma_{2j}^y \sigma_{2j+1}^y, \hspace{1cm} \gamma_{2j} =
\sigma_{2j}^x \sigma_{2j+1}^x,  \label{lv2}
\end{eqnarray}
with $j=1, \cdots,M$. These variables also satisfy the conditions given by
Eq.~(\ref{linkcond}) and the algebra given by Eqs.~(\ref{linkalg1}) and~(\ref%
{linkalg2}). Moreover, we observe that the constraints obeyed now by the
link variables with PBC adopted read 
\begin{equation}
\prod_{j=1}^{M} \eta_{2j-1}\gamma_{2j} = \prod_{j=1}^{M}
\gamma_{2j-1}\eta_{2j} = \mathbb{1}.  \label{cat2}
\end{equation}
By writing $H_{XXZ}$ in terms of $\{\eta_j, \gamma_j\}$ we obtain the same
expression as in Eq.~(\ref{at2}). Therefore, an equivalence between $H_{XXZ}$
and $H_{AT}$ can be achieved as long as the constraints given be Eqs.~(\ref%
{cat}) and~(\ref{cat2}) can be made compatible between each other. In this
direction, we can show that, despite displaying rather different ground
state vectors, the XXZ and Ashkin-Teller models, as defined by Eqs.~(\ref%
{xxz}) and~(\ref{at}), exhibit the same ground state energy. More
specifically, with PBC adopted, it can be shown~\cite{Alcaraz:88} that the
energies in the sector $Q=0$ of $H_{AT}$, which contains the ground state,
will occur in the spectrum of $H_{XXZ}$. Indeed, observe first that
eigenstates of $H_{AT}$ into the sector $Q=0$ are characterized by 
\begin{equation}
\prod_{j=1}^{M} \sigma_j^x = \mathbb{1}, \,\,\,\,\,\, \prod_{j=1}^{M} \tau_j^x
= \mathbb{1},  \label{cond-sym-at}
\end{equation}
which is a consequence of Eq.~(\ref{parity-at}). Then, by inserting Eq.~(\ref%
{cond-sym-at}) into Eq.~(\ref{cat}), we exactly obtain the constraint given
by Eq.~(\ref{cat2}). Therefore, all the energy levels belonging to $Q=0$
also appear in $H_{XXZ}$. Conversely, we can show here that the ground state energy of
the XXZ model is also contained into the spectrum of $H_{AT}$. Indeed, 
besides the U(1) symmetry given by Eq.~(\ref{u1}), $H_{XXZ}$
is also invariant under $Z(2)$ transformations, namely, it commutes with the parity operators 
\begin{equation}
Q_\alpha = \prod_{j=1}^{2M} \sigma_j^\alpha,
\end{equation}
with $\alpha = x,y,z$. In particular, the ground state of $H_{XXZ}$ is into
the sector of eigenstates $\{|\psi\rangle\}$ such that 
\begin{equation}
Q_x |\psi\rangle = + |\psi\rangle, \,\,\,\,\,\, Q_y |\psi\rangle = +
|\psi\rangle.  \label{xxz-parity}
\end{equation}
Then, by using Eq.~(\ref{xxz-parity}) into Eq.~(\ref{cat2}), we exactly
obtain the constraint given by Eq.~(\ref{cat}). Therefore, all the energy
levels belonging to the sector $Q_x = +1, Q_y = +1$ also
appear in $H_{AT}$. Hence, we can conclude that the ground state energy $E_0
(\Delta, \beta)$ of both models are identical for any $\Delta$ and $\beta$.
Remarkably, this equivalence can be extended to the whole spectrum if open
ends are adopted~\cite{Alcaraz:88}. Bearing in mind that $H_{XXZ}$ and $H_{AT}$ have the same
ground state energy, they will manifest the same quantum phase diagram as
the anisotropy $\Delta$ and the staggering parameter $\beta$ are varied.
Naturally, the character of the quantum phase itself depends on the
particular model, since it is associated with the properties of the ground
state vector. 
\begin{figure}[ht]
\centering {\includegraphics[angle=0,scale=0.35]{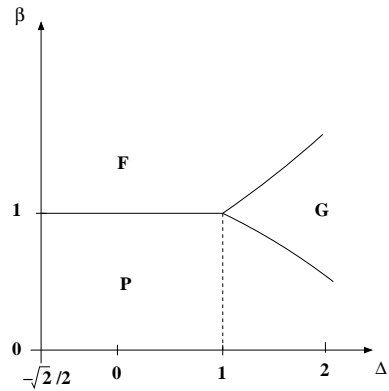}}
\caption{(Color online) Quantum phase diagram of Ashkin-Teller and staggered
XXZ models. 
The energy scale $J$ is set to one.}
\label{f2}
\end{figure}
The quantum phase diagram for $H_{AT}$ and $H_{XXZ}$ is sketched in Fig.~\ref%
{f2}. The quantum phases for the Ashkin-Teller model can be described as the following: 
the ferromagnetic phase (F), where $%
\langle \sigma^z_j\rangle = \langle \tau^z_j\rangle \ne 0$, the paramagnetic
phase (P), where $\langle \sigma^z_j\rangle = \langle \tau^z_j\rangle = 0$,
and the partially ordered phase (G), where $\langle \sigma^z_j
\tau^z_j\rangle \ne 0$. Since our main focus is the investigation of the
infinite-order point $\beta = 1$ and $\Delta = 1$, we will
concentrate on the diagram starting on point $\Delta = -\sqrt{2}/2$, where
we have the beggining of a critical line of continuous QPTs.


\section{Pairwise entanglement}


Let us first analyze the behavior of entanglement for pairs of spins in the
XXZ and Ashkin-Teller models. To this aim, we quantify entanglement by
employing the negativity~\cite{Vidal:02,Verstraete:01}, which is given by 
\begin{equation}
\mathcal{N}(\rho ^{ij})=2\,\max (0,-\min_{\alpha }(\lambda _{\alpha }^{ij})),
\label{neg}
\end{equation}
where $\lambda _{\alpha }^{ij}$ are the eigenvalues of the partial transpose 
$\rho ^{ij,T_{A}}$ of the two-spin density operator $\rho ^{ij}$, defined as 
$\left\langle \alpha \beta \right\vert \rho ^{T_{A}}\left\vert \gamma \delta
\right\rangle =\left\langle \gamma \beta \right\vert \rho \left\vert \alpha
\delta \right\rangle $. For the (non-staggered) XXZ model, pairwise entanglement 
between nearest-neighbors in the ground state of the chain has been evaluated in 
several previous works~\cite{Gu:03,Chen:06,Gu:07,Alcaraz:08}, with entanglement 
measured either by negativity or concurrence~\cite{Wootters:98}. For the specific 
case of ground state entanglement in the XXZ model, negativity and concurrence turn out to be 
identical~\cite{Wang:02,Alcaraz:08}. In particular, at the infinite-order QCP 
$\Delta = 1$, it has been shown that the negativity achieves a maximum given by 
$0.386$. Indeed, this maximum is rather robust, appearing not only in
the ground state but also for all conformal towers associated with $H_{XXZ}$
formed by an infinite number of excited states~\cite{Alcaraz:08}.

\begin{figure}[th]
\centering {\includegraphics[angle=0,scale=0.26]{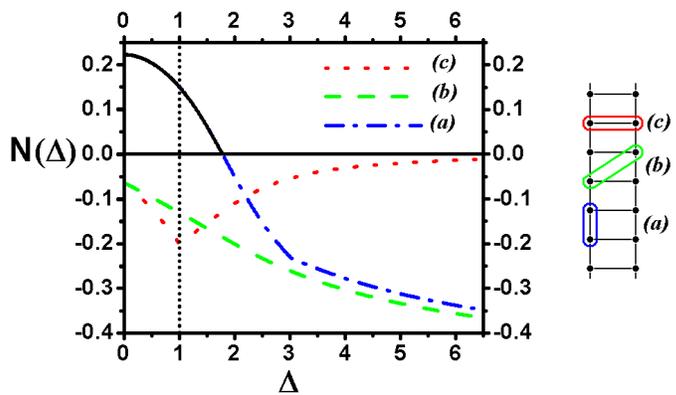}}
\caption{(Color online) Negativity $\mathcal{N}$ (solid black line) and DSB $%
\Lambda$ for the pairs (a) $\protect\sigma_j-\protect\sigma%
_{j+1}$ and $\protect\tau_j-\protect\tau_{j+1}$,
(b) $\protect\sigma_j-\protect\tau_{j+1}$, and (c) $\protect\sigma_j-\protect\tau_j$ 
in the Ashkin-Teller chain with 20 spins. Note that the only nearest spin pairs with nonvanishing
entanglement are given by curve (a). The results are plotted for $\beta = 1$.}
\label{f3}
\end{figure}

\begin{figure}[th]
\centering {\includegraphics[angle=0,scale=0.28]{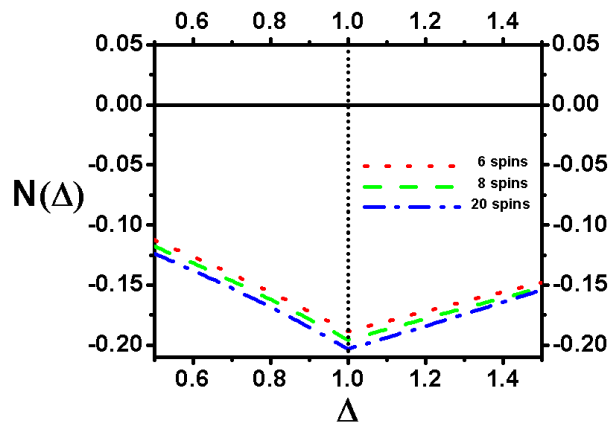}}
\caption{(Color online) Negativity $\mathcal{N}$ (solid black line) and DSB $%
\Lambda$ for a frontal pair $\protect\sigma_j-\protect\tau_j$ for chains with $%
6$, $8$, and $20$ spins (from top to bottom) as a function of $\Delta$. Note
that the operation $\max$ hides the nonanalyticity exhibited by $\Lambda$. 
The results are plotted for $\beta = 1$.}
\label{f4}
\end{figure}

Concerning the Ashkin-Teller model, we have two different spins at each
site, given by the Pauli operators $\sigma$ and $\tau$. We will consider here
entanglement between pairs $\sigma_j-\tau_{j+1}$, $\sigma_j-\sigma_{j+1}$, and $%
\tau_j-\tau_{j+1}$ at nearest neighbour sites as well as pairs $\sigma_j-\tau_j
$ at the same site (frontal pairs). Since the Hamiltonian is symmetric by an
interchange $\sigma \leftrightarrow \tau$, both $\sigma_j-\sigma_{j+1}$ and $%
\tau_j-\tau_{j+1}$ exhibit the same entanglement properties. The negativity
for these pairs is rather different of that for nearest neighbor spins in the XXZ
chain, displaying no signature (e.g., a maximum) of the infinite-order QPT.
This result is plotted in Fig.~\ref{f3}(a). 
For the spins $\sigma_j-\tau_j$ and $\sigma_j-\tau_{j+1}$, we can show that the
negativity is vanishing for any $\Delta$, as given by Figs.~\ref{f3}(b) and~\ref{f3}(c). 
Therefore, entanglement for these pairs do not exhibit any indication of the quantum critical 
behavior. However, an indication of the infinite-order QPT can be obtained by looking
at a closely related quantity, which is the distance of the separability
boundary (DSB), denoted by $\Lambda (\rho ^{ij})$. By following Ref.~\cite%
{Yu:07}, we can define $\Lambda (\rho ^{ij})$ from Eq.~(\ref{neg}) through 
\begin{equation}
\Lambda(\rho ^{ij}) = - 2 \min_{\alpha }(\lambda _{\alpha }^{ij}).
\label{dsb}
\end{equation}
Note that $\Lambda > 0$ implies an entangled state with $\Lambda (\rho
^{ij}) = \mathcal{N}(\rho ^{ij})$. It can be shown that $\Lambda = 0$ for 
\textit{pure} separable states while $\Lambda < 0$ for \textit{mixed} separable
states. Results for $\Lambda$ in a chain with 20 spins are shown in Fig.~\ref{f3}(a), 
~\ref{f3}(b), and~\ref{f3}(c). Note that the distance $\Lambda$ for the
frontal pair of spins presents a cusp at the infinite-order QCP. This cusp
is actually hidden by the $\max$ operation and is kept unchanged for larger
chains, as shown in Fig.~\ref{f4}. 
In order to understand the nonanalyticity of $\Lambda$ in $\Delta = 1$, let
us consider the density matrix for $\sigma_j-\tau_j$, which is given by 
\begin{equation}
\rho^{j,j}=\left( 
\begin{array}{cccc}
u (\Delta) & 0 & 0 & 0 \\ 
0 & v (\Delta) & 0 & 0 \\ 
0 & 0 & v (\Delta) & 0 \\ 
0 & 0 & 0 & w (\Delta)%
\end{array}%
\right) ,
\end{equation}
where $u(\Delta) = \frac{1}{4}+\frac{1}{2}m\left( \Delta \right) +\frac{1}{4}%
G\left( \Delta \right)$, $v (\Delta)= \frac{1}{4}-\frac{1}{4}G\left( \Delta
\right)$, and $w (\Delta)= \frac{1}{4}-\frac{1}{2}m\left( \Delta \right) +%
\frac{1}{4}G\left( \Delta \right)$, with $m\left( \Delta \right) $ denoting
the magnetization density in the $x$ direction, namely, $m\left( \Delta \right) =
\left\langle \sigma_{k}^{x}\right\rangle= \left\langle
\tau_{k}^{x}\right\rangle$ and $G\left( \Delta \right) = \left\langle \sigma_{k}^{x}\tau_{k}^{x}\right\rangle$.
Since the density matrix is diagonal, the partial transposition keeps the
same eigenvalues as $\rho^{j,j}$. Then, it can be numerically shown that the
DSB reads 
\begin{eqnarray}
\Lambda =\left\{ 
\begin{array}{l}
-\frac{1}{2}+m\left( \Delta \right) -\frac{1}{2}G\left( \Delta \right)
\qquad \text{if }\Delta \leq 1, \\ 
-\frac{1}{2}+\frac{1}{2}G\left( \Delta \right) \qquad \qquad \qquad \text{if 
}\Delta >1.%
\end{array}%
\right.
\end{eqnarray}
Therefore, there is a change in the behavior of $\Lambda$ exactly at $\Delta
= 1$, which makes the DSB a useful pairwise quantity capable of identifying
the infinite-order QCP in the Ashkin-Teller model.


\section{Block entanglement}


\label{be}

As we have seen, in contrast with the XXZ model, pairwise entanglement alone
(with no DSB suplementary analysis) fails to identify the infinite-order QCP
in the Ashkin-Teller model. In this section, we will show that bipartite
block entanglement as measured by the von Neumann entropy is able to provide
a unique description of entanglement for both the XXZ and Ashkin-Teller
models. Given a quantum system in a pure state $|\psi\rangle$ and a
bipartition of the system into two blocks $A$ and $B$, entanglement between $%
A$ and $B$ can be measured by the von Neumann entropy $\mathcal{S}$ of the
reduced density matrix of either of blocks, i.e., 
\begin{equation}
\mathcal{S}=-\text{Tr} \left( \rho_A \log_2 \rho_A \right) = -\text{Tr}
\left( \rho_B \log_2 \rho_B \right),  \label{vonNeumann}
\end{equation}
where $\rho_A=\text{Tr}_B \rho$ and $\rho_B = \text{Tr}_A \rho$ denote the
reduced density matrices of blocks $A$ and $B$, respectively, with $%
\rho=|\psi\rangle\langle \psi|$.

\begin{figure}[th]
\centering {\includegraphics[angle=0,scale=0.25]{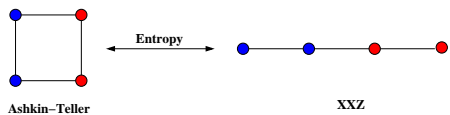}}
\caption{(Color online) Map of spins to obtain the equivalence of the
entanglement entropies for the Ashkin-Teller and XXZ models.}
\label{f5}
\end{figure}

Concerning the XXZ model, it has been shown that local extremes of the von
Neumann entropy are able to identify the quantum phase diagram~\cite{Chen:06}. 
In the case of the Ashkin-Teller model, we will show here that the
entanglement entropy can also identify the infinite-order QPT by a local
extreme, displaying a superior behavior in comparison with pairwise
entanglement. Moreover, for a suitable choice of blocks, we can also show
that the entanglement entropy is the same as that of the XXZ
model. Indeed, a sublattice composed by a {\it contiguous} block of spins in the
XXZ chain will exhibit the same von Neumann entropy as a set of {\it contiguous}
frontal pairs of spins in the Ashkin-Teller model. This is analytically
worked out in the Appendix for a two-spin block and numerically checked for
larger blocks. The map of equivalent entropies is skechted in Fig.~\ref{f5}.

\begin{figure}[th]
\centering {\includegraphics[angle=0,scale=0.24]{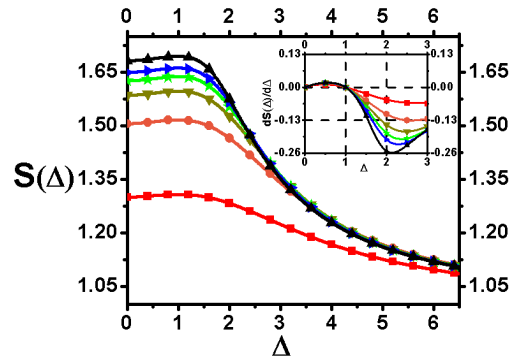}}
\caption{(Color online) Block entanglement $S(\Delta)$ as a function of the
anisotropy $\Delta$ for a single frontal pair of spins. From bottom to top, 
the chain lenghts are $6$, $8$, $10$, $12$, $14$, and $20$ spins, respectively. 
Inset: the derivatives of $S(\Delta)$ vanish at $\Delta = 1$. }
\label{f6}
\end{figure}

\begin{figure}[th]
\centering {\includegraphics[angle=0,scale=0.21]{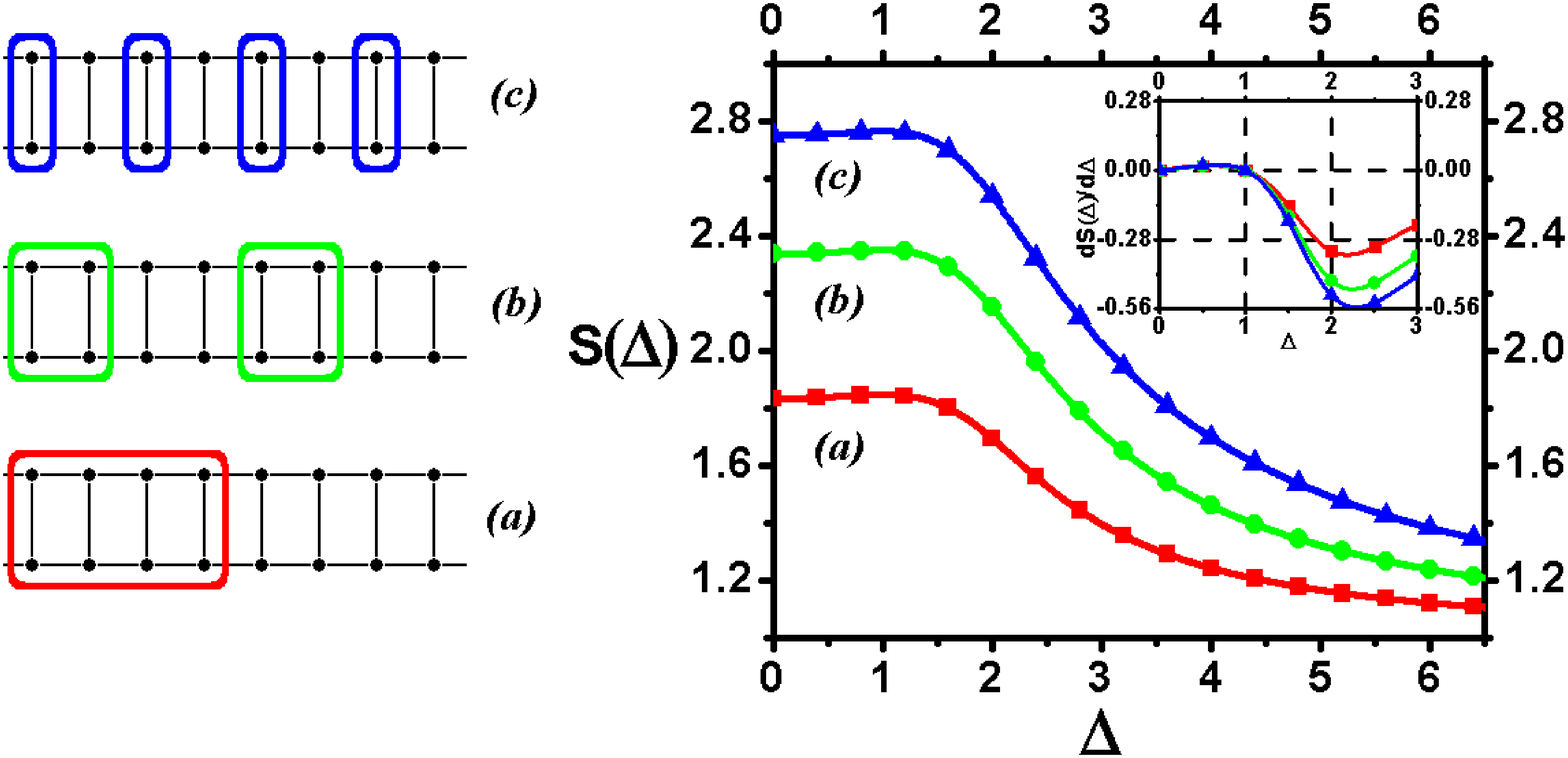}}
\caption{(Color online) Block entanglement $S(\Delta)$ as a function of the
anisotropy $\Delta$ for $L=4$ sites in different configurations in a chain 
with $N=20$ spins. Inset: derivatives of $S(\Delta)$, all of them vanishing 
at $\Delta = 1$.}
\label{f7}
\end{figure}

Note that the map above refers to the von Neumann entropy only. Naturally,
it is not applicable to obtain equivalences in the case of other
physical quantities, e.g., pairwise entanglement. As we have seen, no
equivalence can be obtained in that case. In order to investigate the
characterization of the QPT by using bipartite block entanglement, let us
first consider the situation $\beta = 1$ (non-staggered model). Then, by
performing exact diagonalization, we plot in Fig.~\ref{f6} the von Neumann
entropy $S(\Delta)$ for a single frontal pair of spins in a chain with
different lengths as a function of the anisotropy $\Delta$. Note that,
independently of the chain size, $S(\Delta)$ presents a maximum exactly at 
$\Delta =1$ (see Inset). We can also show that the maximum at the critical
point is robust against the increase of the size of the blocks as well as
the choice of the sublattices. This is particularly shown in Fig.~\ref{f7},
where it is exhibited the entropy in the Ashkin-Teller model for different 
choices of sublattices in a chain with 20 spins. Note that the entropy gets 
larger as we increase the number of bonds between the sublattices, i.e., 
$S(\Delta)$ increases according to the direction {\it red} [curve (a)] to 
{\it blue} [curve (c)] of Fig.~\ref{f7}. 
Again, a maximum is observed at $\Delta = 1$ (see inset). As mentioned before, 
entanglement in the Ashkin-Teller model will match entanglement in the XXZ chain 
for contiguous blocks of spins (e.g., the {\it red} plot [curve (a)] of Fig.~\ref{f7}).


\begin{figure}[th]
\centering {\includegraphics[angle=0,scale=0.28]{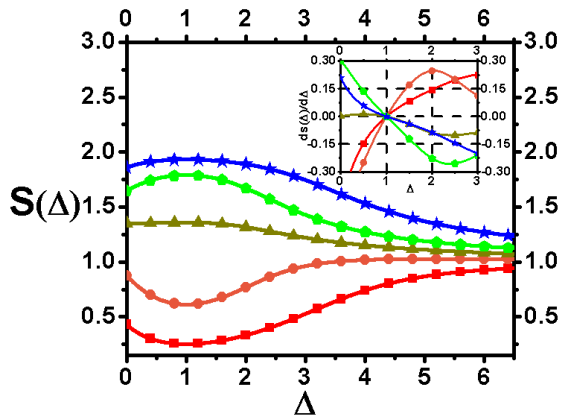}}
\caption{(Color online) Entanglement entropy for a frontal pair $\protect\sigma%
_j-\protect\tau_j$ in the Ashkin-Teller chain with 20 spins as a function of 
$\Delta$ for different values of the staggering couplings $\protect\beta$.
From bottom to top, the figures are plotted for $\protect\beta=\frac{1}{2}$, 
$\frac{3}{4}$, $1$, $\frac{5}{4}$, and $\frac{7}{4}$,
respectively. Inset: the derivatives of $S(\Delta)$ vanish at $\Delta = 1$
and indicate a pronounced maximum (or minimum) around the second-order QPT.}
\label{f8}
\end{figure}

Hence, the von Neumann entropy provides a completely equivalent description of
block entanglement in the XXZ and Ashkin-Teller models, being able to
characterize the QPT by a maximum at $\Delta = 1$. We shall now analyze
the case of staggering parameters $\beta \ne 1$. Let us consider the
entanglement between a frontal pair and all the rest of the chain in the
Ashkin-Teller model. The resulting entropy for a chain with $20$ spins is
plotted in Fig.~\ref{f8} as a function of $\Delta$ for several values of the
staggering parameter $\beta$. 

\begin{figure}[th]
\centering {\includegraphics[angle=0,scale=0.28]{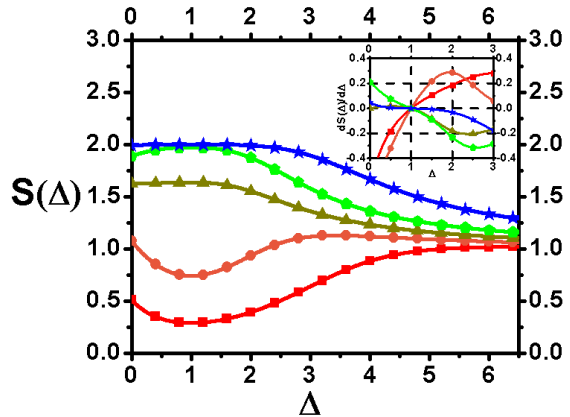}}
\caption{(Color online) Entanglement entropy for a quartet in the
Ashkin-Teller chain with 20 spins as a function of $\Delta$ for different
values of the staggering couplings $\protect\beta$. From bottom to top, the
figures are plotted for $\protect\beta=\frac{1}{2}$, $\frac{3}{4}$, $1$, $%
\frac{5}{4}$, and $\frac{7}{4}$, respectively. Inset: the
derivatives of $S(\Delta)$ vanish at $\Delta = 1$ and indicate a pronounced
maximum (or minimum) around the second-order QPT. These maxima are expected to 
scale as we increase the size of the chain.}
\label{f9}
\end{figure}

Note that for $\beta \ge 1$ the von Neumann entropy displays a maximum at
$\Delta = 1$ while for $\beta < 1$ it is characterized by a minimum at 
$\Delta = 1$. The change in the concavity of $S(\Delta)$ at $\beta = 1$ 
is rather robust, being independent of the size of the chain. However, 
by looking at the quantum phase diagram at Fig.~\ref{f2}, we can see that, 
for $\beta \ne 1$, there is no infinite-order QPT at $\Delta = 1$. 
Therefore, we are providing here an example of a local extreme of entanglement 
that is not associated with a QPT. Nontrivial staggering is then able to 
introduce a phenomenon that is absent in the standard case. Nevertheless, 
observe that the concavity of the curve can characterize the paramagnetic-
ferromagnetic QPT, namely, the von Neumann entropy always displays a maximum 
at the ferromagnetic (F) case while a minimum is found in the paramagnetic (P) 
region. Therefore, the concavity of $S(\Delta)$ provides a necessary condition 
(that is also sufficient for $\Delta \le 1$) to determine the phases F and P. 
Moreover, note that entanglement also detects both the P-G and F-G QPTs of 
Fig.~\ref{f2}, given by horizontal crossings at the phase diagram for a fixed 
$\beta$. Indeed, these P-G and F-G crossings are second-order QPTs, with the 
first derivative of entanglement getting a pronounced maximum (or minimum) at 
the QPT. This is exhibited in the Inset of Fig.~\ref{f8}. The behavior above 
is also kept if we increase the size of the block. Indeed, we plot in 
Fig.~\ref{f9} the entanglement between a quartet composed by two frontal spins 
and all the rest of the chain, which exhibits the same pattern of maxima and 
minima for the entropy as in the case of single frontal spin pairs. 

\begin{figure}[th]
\centering {\includegraphics[angle=0,scale=0.28]{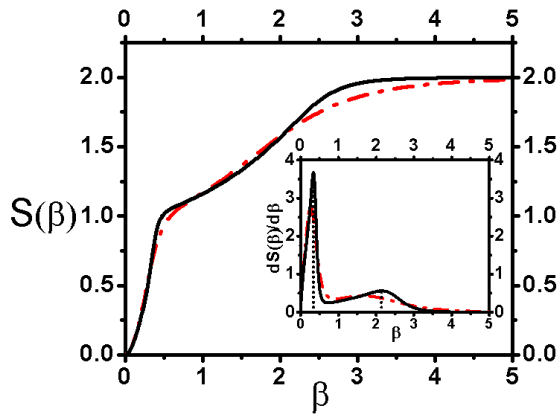}}
\caption{(Color online) Entanglement entropy for a quartet in the
Ashkin-Teller chain with 8 (dashed red curve) and 20 spins (solid black curve) as a function of $\beta$ 
for $\Delta = 5$. Inset: the first derivative of $S(\beta)$ displays pronounced maxima at 
$\beta \approx 0.337$ and $\beta \approx 2.14$, which are finite-size precursors of the second-order QPTs. 
These maxima are expected to scale as we increase the size of the chain.}
\label{f10}
\end{figure}

We observe that the pronunciation of the maximum or the minimum of the
derivative of entanglement at a second-order QPT is expected to evolve to a nonanalyticity as we
increase the size the chain~\cite{Wu:04,Wu:06}. This is indicated in Fig.~\ref{f10}, where the entropy as a function 
of $\beta$ reveals the second-order QPTs by vertical crossings at the phase diagram given by Fig.~\ref{f2} for a fixed 
$\Delta$.

Note from Fig.~\ref{f10} that, for any size of the chain, entanglement entropy for a quartet tends to 2 as 
$\beta \rightarrow \infty$. This can be understood as a consequence of the map shown in Fig.~\ref{f5},  
where we replace the entropy of the quartet by the entropy of four contiguous spins in the XXZ chain. 
Indeed, in the limit $\beta \rightarrow \infty$, the ground state of the Hamiltonian~(\ref{xxz})  
is given by a set decoupled dimers, whose density operator is $\rho = \rho_{2,3} \otimes \rho_{4,5} \otimes 
\cdots \rho_{2M,1}$, with $\rho_{2j,2j+1} = (1/4)(\mathbb{1}_{2j}\otimes\mathbb{1}_{2j+1} + 
\sigma^x_{2j}\otimes\sigma^x_{2j+1} + \sigma^y_{2j}\otimes\sigma^y_{2j+1} - \sigma^z_{2j}\otimes\sigma^z_{2j+1})$. 
With no loss of generalility, by analyzing the quartet $\rho_{1,2,3,4}$, we obtain that 
$\rho_{1,2,3,4} = (1/4) (\mathbb{1}_1\otimes\rho_{2,3}\otimes\mathbb{1}_4)$. Then it follows that $S(\rho_{1,2,3,4}) = 2$, which 
comes from the contribution of the spin singlets $(2M,1)$ and $(4,5)$.

\section{Conclusions}

In summary, we have presented a detailed analysis of entanglement into two
closely related models, namely, the staggered XXZ and Ashkin-Teller models. In
particular, the behavior of the entanglement properties in both models has
been considered at the infinite-order QCP, where we have shown that pairwise
entanglement in the Ashkin-Teller model fails in the characterization of the
QPT (with no DSB supplementary analysis). However, the von Neumann entropy
can offer a unified description of quantum criticality, with this
equivalence achieved for contiguous blocks of spins. Moreover, we have also
shown that the pattern of local extremes at the infinite-order QCP has been
obtained by using bipartite block entropy for different choices of sublattices, 
even though local extremes also appear in regions where no QPTs occur. This result 
establishes that the maxima and minima of entanglement entropy {\it per se} constitute, in 
the absence of supplentary analysis, {\it insufficient} conditions to ensure the 
existence of an infinite-order QPT.

The local extremes parttern of entanglement for infinite-order QPTs has been
observed for a variety of different systems, despite no analytical
derivation of this behavior is available. This is in contrast with
first-order and continuous QPTs, where the relationship between QPTs and
nonanalytical behavior of entanglement has been generically understood for
finite dimensional bipartite~\cite{Wu:04} and multipartite~\cite{Oliveira:06}
systems as well as for continuous variable models~\cite{Rieper:09}. We have
provided here new indications that local extremes provide a typical but not sufficient 
characterization of infinite-order QPTs, which may provide a hint for a future derivation 
of this phenomenon in general grounds.

\subsection*{Acknowledgments}

We thank Prof. F. C. Alcaraz for helpful discussions. 
The authors acknowledge financial support from the Brazilian funding agencies MCT/CNPq and FAPERJ.
This work was performed as part of the Brazilian National Institute for Science and Technology of Quantum 
Information (INCT-IQ).


\section*{Appendix}


Let us show in this appendix that the reduced density operator $\rho^{AT}_j$
for a frontal spin pair at site $j$ in the Ashkin-Teller is equivalent to a
nearest-neighbor two-spin reduced density operator $\rho^{XXZ}_{j,j+1}$ in
the XXZ chain. Starting from the Ashkin-Teller model, $\rho^{AT}$ reads 
\begin{equation}
\rho^{AT}_j = \frac{1}{4}\left( \mathbb{1}_4 + u \,\sigma_{j}^{x} +u \,
\tau_{j}^{x} + v \,\sigma _{j}^{x} \tau_{j}^{x}\right),  \label{rho2at}
\end{equation}
where $\mathbb{1}_4$ is the 4-dimensional identity operator, $u = \langle
\sigma_j^{x} \rangle = \langle \tau_j^{x} \rangle$, and $v = \langle
\sigma^x_j \tau_j^{x} \rangle$. By rewriting Eq.~(\ref{rho2at}) in terms of
the link variables given by Eq. (\ref{lv}) we obtain 
\begin{equation}
\rho^{AT}_j = \frac{1}{4}\left( \mathbb{1}_4 + u \, \eta_{2j-1} + u\,
\gamma_{2j-1} + v\, \eta_{2j-1} \gamma_{2j-1}\right),  \label{rho2atlv}
\end{equation}
with $u = \langle \eta_{2j-1} \rangle = \langle \gamma_{2j-1} \rangle$ and $%
v = \langle \eta_{2j-1} \gamma_{2j-1} \rangle$. Now, let us turn to the XXZ
chain, whose two-spin reduced density operator reads 
\begin{equation}
\rho^{XXZ}_{j,j+1} =\frac{1}{4}\left( \mathbb{1}_4 + p \,\sigma
_{j}^{x}\sigma _{j+1}^{x} + p\, \sigma _{j}^{y}\sigma _{j+1}^{y}+ q\,\sigma
_{j}^{z}\sigma _{j+1}^{z}\right),
\end{equation}
where $p = \langle \sigma _{j}^{x}\sigma _{j+1}^{x} \rangle = \langle \sigma
_{j}^{y}\sigma _{j+1}^{y} \rangle$, and $q = \langle \sigma _{j}^{z}\sigma
_{j+1}^{z} \rangle$. In terms of the link variables given by Eq.~(\ref{lv2}%
), we obtain 
\begin{equation}
\rho^{XXZ}_j = \frac{1}{4}\left( \mathbb{1}_4 + p \, \eta_{2j-1} + p\,
\gamma_{2j-1} - q\, \eta_{2j-1} \gamma_{2j-1}\right),  \label{rho2xxzlv}
\end{equation}
with $p = \langle \eta_{2j-1} \rangle = \langle \gamma_{2j-1} \rangle$ and $%
q = - \langle \eta_{2j-1} \gamma_{2j-1} \rangle$. Hence, since $u=p$ and $%
v=-q$, we have that Eq.~(\ref{rho2atlv}) is equivalent to Eq.~(\ref%
{rho2xxzlv}). Therefore, the eigenvalues of $\rho^{AT}_j$ are identical to
the eigenvalues of $\rho^{XXZ}_{j,j+1}$, which means that the von Neumann
entropy are the same in both cases.

\end{document}